\documentclass{article}

\usepackage{microtype}
\usepackage{graphicx}
\usepackage{subfigure}
\usepackage{booktabs} 

\usepackage{hyperref}

\usepackage[accepted]{icml2025}

\usepackage{amsmath}
\usepackage{amssymb}
\usepackage{mathtools}
\usepackage{amsthm}

\usepackage[capitalize,noabbrev]{cleveref}

\theoremstyle{plain}
\newtheorem{theorem}{Theorem}[section]

\theoremstyle{definition}
\newtheorem{definition}[theorem]{Definition}

\theoremstyle{remark}

\usepackage[textsize=tiny]{todonotes}

\icmltitlerunning{}

\begin{document}

\twocolumn[
\icmltitle{The Cognitive Foundations of Economic Exchange: A Modular Framework Grounded in Behavioral Evidence}




\begin{icmlauthorlist}
\icmlauthor{Egil Diau}{NTU}
\end{icmlauthorlist}

\icmlaffiliation{NTU}{Department of Computer Science, Taipei, Taiwan}

\icmlcorrespondingauthor{Egil Diau}{egil158@gmail.com}

\icmlkeywords{Machine Learning, ICML}

\vskip 0.3in
]



\printAffiliationsAndNotice{} 

\begin{abstract}
The origins of economic behavior remain unresolved—not only in the social sciences but also in AI, where dominant theories often rely on predefined incentives or institutional assumptions. Contrary to the longstanding myth of barter as the foundation of exchange, converging evidence from early human societies suggests that reciprocity—not barter—was the foundational economic logic, enabling communities to sustain exchange and social cohesion long before formal markets emerged. Yet despite its centrality, reciprocity lacks a simulateable and cognitively grounded account. Here, we introduce a minimal behavioral framework based on three empirically supported cognitive primitives—individual recognition, reciprocal credence, and cost--return sensitivity—that enable agents to participate in and sustain reciprocal exchange, laying the foundation for scalable economic behavior. These mechanisms scaffold the emergence of cooperation, proto-economic exchange, and institutional structure from the bottom up. By bridging insights from primatology, developmental psychology, and economic anthropology, this framework offers a unified substrate for modeling trust, coordination, and economic behavior in both human and artificial systems. For an interactive visualization of the framework, see: \url{https://egil158.github.io/cogfoundations-econ/}
\end{abstract}

\begin{figure}[htbp]
  \centering
  \includegraphics[width=0.45\textwidth]{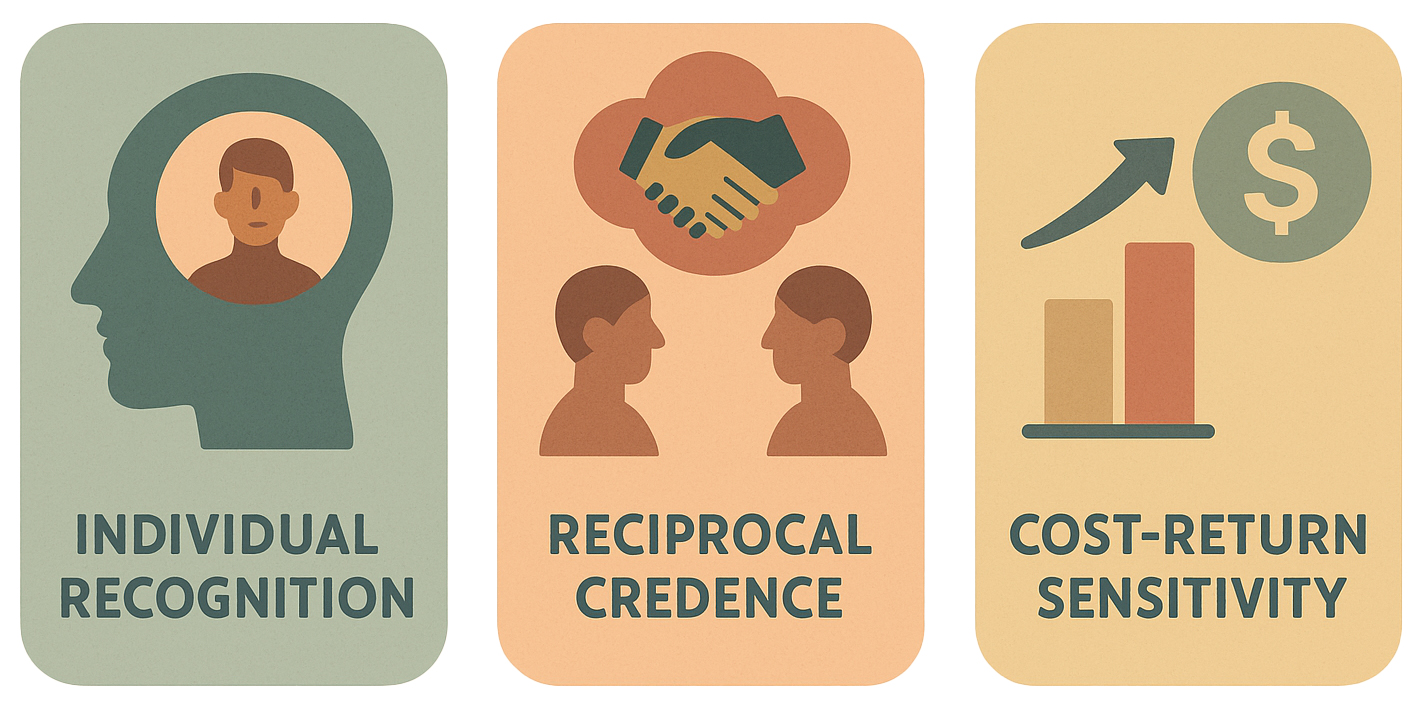}
  \caption{Three Core Cognitive Mechanisms—Individual Recognition, Reciprocal Credence, and Cost–Return Sensitivity—as Behavioral Primitives for Simulating Reciprocal Exchange in Artificial Agents.}
  \label{fig:main}
\end{figure}

\section{Introduction}
\label{submission}

The standard origin story of economics begins with barter: individuals trading goods directly, with money and markets emerging to reduce friction. Yet this narrative is a myth. Ethnographic and historical research finds little evidence for barter as a primary mode of early exchange \cite{sahlins2013stone, mauss2024gift, malinowski2013argonauts}. Instead, early human societies were structured around \textit{reciprocity}—a temporally extended system of giving, receiving, and returning, embedded in social relationships.

Surprisingly, most economic models still begin where barter leaves off: with institutions, contracts, and payoff matrices. These frameworks assume—but do not explain—the behavioral substrate that makes cooperation possible in the first place. As a result, we lack simulateable accounts of how economic behavior emerges from interaction among agents.

This blind spot extends to broader AI + society research. Core concepts like “trust,” “value,” and “cooperation” are often used without precise definitions or behavioral grounding. Rather than being modeled as well-defined, simulateable mechanisms, they are treated as vague abstractions—obscuring rather than explaining the dynamics of social interaction.

In this paper, we argue that the foundations of economic exchange—and by extension, scalable cooperation—emerge not from barter or institutional design, but from \textit{reciprocity}: a structured pattern of interaction built on three cognitively minimal mechanisms—
\textit{individual recognition}, \textit{reciprocal credence}, and \textit{cost–return sensitivity}. These mechanisms, observed across humans and nonhuman primates, support partner-specific cooperation without relying on symbolic trust, contracts, or centralized enforcement.

We formalize these mechanisms as simulateable behavioral primitives, providing a bottom-up framework grounded in cognitive and behavioral evidence. By integrating insights from anthropology, developmental psychology, and economic anthropology, our approach offers a biologically grounded alternative to institution-first models of cooperation.

\paragraph{Our contribution.} We provide a minimal theoretical framework grounded in behaviorally observable mechanisms. This framework yields four core contributions:

\begin{itemize}
    \item Identify three minimal mechanisms—individual recognition, reciprocal credence, and cost–return sensitivity—as simulateable behavioral primitives for modeling the emergence of economic exchange;
    \item Synthesize evidence from primate behavior, infant cognition, and economic anthropology to support these mechanisms across species and developmental stages;
    \item Reframe “trust” as a scalar, simulateable expectation—reciprocal credence—rather than a moral abstraction;
    \item Reinterpret classic behavioral findings—such as framing effects, loss aversion, and anchoring bias—as consistent expressions of bounded, biologically grounded cognition, rather than anomalies to rational choice theory.
\end{itemize}

Unlike models that treat trust heuristically or assume cooperation as exogenous, our framework offers a minimal, simulateable foundation for how cooperative expectations and exchange structures emerge from interaction. This allows us to reframe the origins of economic systems not as top-down implementations of institutions, but as bottom-up constructions grounded in cognitive behavior.

\paragraph{Ethical Statement} Importantly, this work does not rely on evolutionary explanations. While we draw on behavioral evidence from primates and human infants, our goal is not to claim innate or adaptive origins of exchange. Rather, these cases serve as empirical constraints to identify minimal cognitive mechanisms sufficient for reciprocal behavior. Our account is grounded in behavioral plausibility, not evolutionary teleology.

\section{Related Work}
\subsection{Agent-Based Social Simulation}

Recent work on multi-agent language models has enabled scripted cooperation and task planning \citep{park2023generative, li2023camel}, but these systems rarely model the cognitive mechanisms that support stable social interaction—such as partner tracking, interaction memory, or sensitivity to prior outcomes.

While some frameworks add memory or heuristics, these are typically unconstrained and ad hoc, limiting their ability to simulate reciprocal dynamics over time. Apparent cooperation often stems from prompt bias or hardcoded behavior rather than simulateable social inference.

Our framework addresses this gap by identifying minimal, biologically grounded mechanisms that support dynamic reciprocity, enabling more realistic modeling of social behavior and scalable exchange.

\subsection{Reciprocal Behavior in Nonhuman Primates}

Studies in primatology reveal that species such as chimpanzees and bonobos engage in reciprocal acts across grooming, food sharing, and alliance formation. For instance, de Waal \cite{de1997chimpanzee} documented long-term cooperative relationships maintained through delayed reciprocity, while Brosnan et al. \cite{brosnan2003monkeys} showed sensitivity to fairness and outcome inequity.

Although these findings offer strong evidence of reciprocal behavior, they are seldom linked to broader economic structures or coordination dynamics. The underlying cognitive mechanisms—and how they scale to support complex social systems—remain largely unexplored.

\subsection{Social Exchange Without Markets}

Anthropological work has challenged the classical view that early economies began with barter. Foundational accounts, such as Mauss’s \textit{The Gift} \cite{mauss2024gift} and Sahlins’s \textit{Stone Age Economics} \cite{sahlins2013stone}, emphasize that early exchange was embedded in networks of kinship, obligation, and prestige—rather than driven by market logic or equivalent trade.

These studies highlight the importance of social context and long-term reciprocity, but often leave open the question of what mechanisms make such systems sustainable. Specifically, the cognitive conditions under which deferred, partner-contingent exchange can stabilize across time are rarely formalized.

\begin{figure}[htbp]
  \centering
  \includegraphics[width=0.45\textwidth]{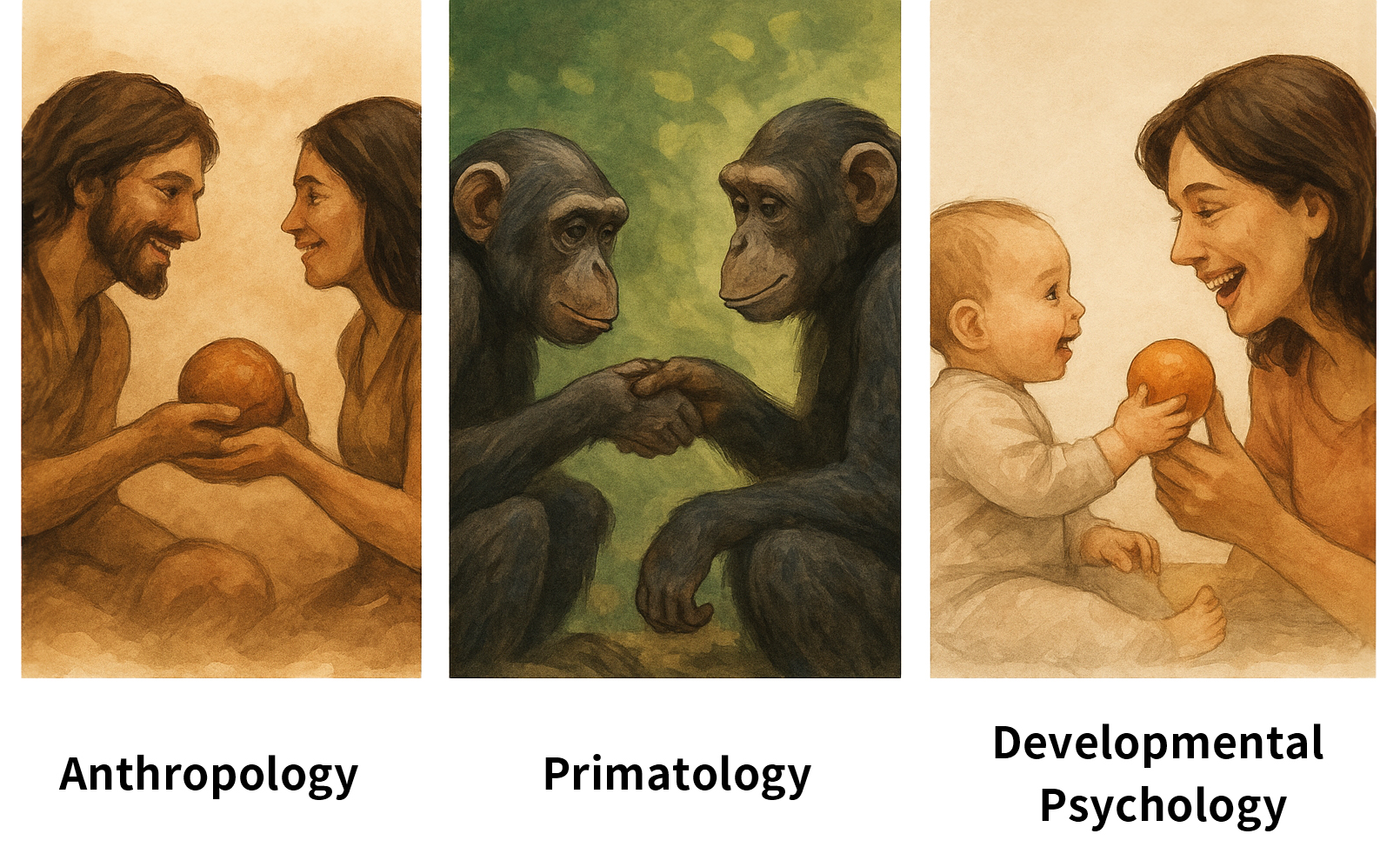}
  \caption{Evidence from anthropology, primatology, and developmental psychology suggests that the behavioral foundations of economic exchange lie in reciprocity—not barter or market logic.}
  \label{fig:source}
\end{figure}

\section{Background: Behavioral Origins of Exchange}
\label{headings}

\subsection{The Myth of Barter and the Reciprocal Foundations of Exchange}

The standard narrative of economic origins begins with barter: the idea that early humans exchanged goods directly, with markets and money emerging later to reduce friction. But this account lacks support in ethnographic and historical records.

As Sahlins argued in \textit{Stone Age Economics} \cite{sahlins2013stone}, early exchange was not based on equivalence but on relational reciprocity: generalized reciprocity involved open-ended giving among kin; balanced reciprocity entailed delayed equivalence among peers; and negative reciprocity reflected opportunism among strangers. Malinowski’s Kula ring \cite{malinowski2013argonauts} exemplifies long-term, prestige-based exchange using non-utilitarian objects, governed by memory, exclusivity, and directional flow. Beneath the ritual lies a cognitively rich structure of tracking, obligation, and reputation. Mauss’s \textit{The Gift} \cite{mauss2024gift} formalized this logic as a triadic obligation—to give, receive, and reciprocate—framing gift exchange as a cognitive and social mechanism for sustaining long-term bonds.

Together, these accounts suggest that early human societies were not built on barter, but on reciprocity: a temporally extended structure of interaction grounded in memory, obligation, and repeated engagement.

\textit{Far from a peripheral exchange strategy, reciprocity was the behavioral infrastructure of early economies—the cognitive and social foundation that enabled resource flow, social cohesion, and relational continuity.}

It sustained cooperation long before the emergence of contracts or currency, and predates markets not only historically, but cognitively—anchored in the minimal behavioral logic we formalize in the next section.

\subsection{The Three Behavioral Mechanisms Enabling Reciprocity}

If reciprocity—not barter—underpins early economic life, its roots must predate formal institutions. This raises a fundamental question: what minimal behavioral and cognitive mechanisms are sufficient to sustain reciprocal exchange?

Nonhuman primates offer a critical comparative lens. In chimpanzees and bonobos, sustained, partner-contingent behaviors—such as food sharing, grooming, and coalition support—exhibit structured patterns of reciprocity and serve as core mechanisms for maintaining group cohesion and social stability \cite{de1997chimpanzee}.

Comparative psychology isolates the minimal cognitive substrates that support such behaviors. By examining species with simpler interaction structures, it avoids cultural and institutional confounds—revealing the baseline capacities sufficient for reciprocity to emerge and stabilize.

To formalize these foundations, we introduce a cognitively grounded framework comprising three simulateable mechanisms:
\begin{itemize}
    \item \textbf{Individual recognition}: identifying and re-engaging specific social partners over time;
    \item \textbf{Reciprocal credence}: an updateable expectation that cooperation will be returned;
    \item \textbf{Cost–return sensitivity}: modulating cooperative behavior in response to payoff asymmetries.
\end{itemize}

We treat these as simulateable behavioral primitives—providing a bottom-up foundation for modeling the emergence of economic exchange, without presupposing institutions, markets, or symbolic trust.

\begin{figure}[htbp]
  \centering
  \includegraphics[width=0.45\textwidth]{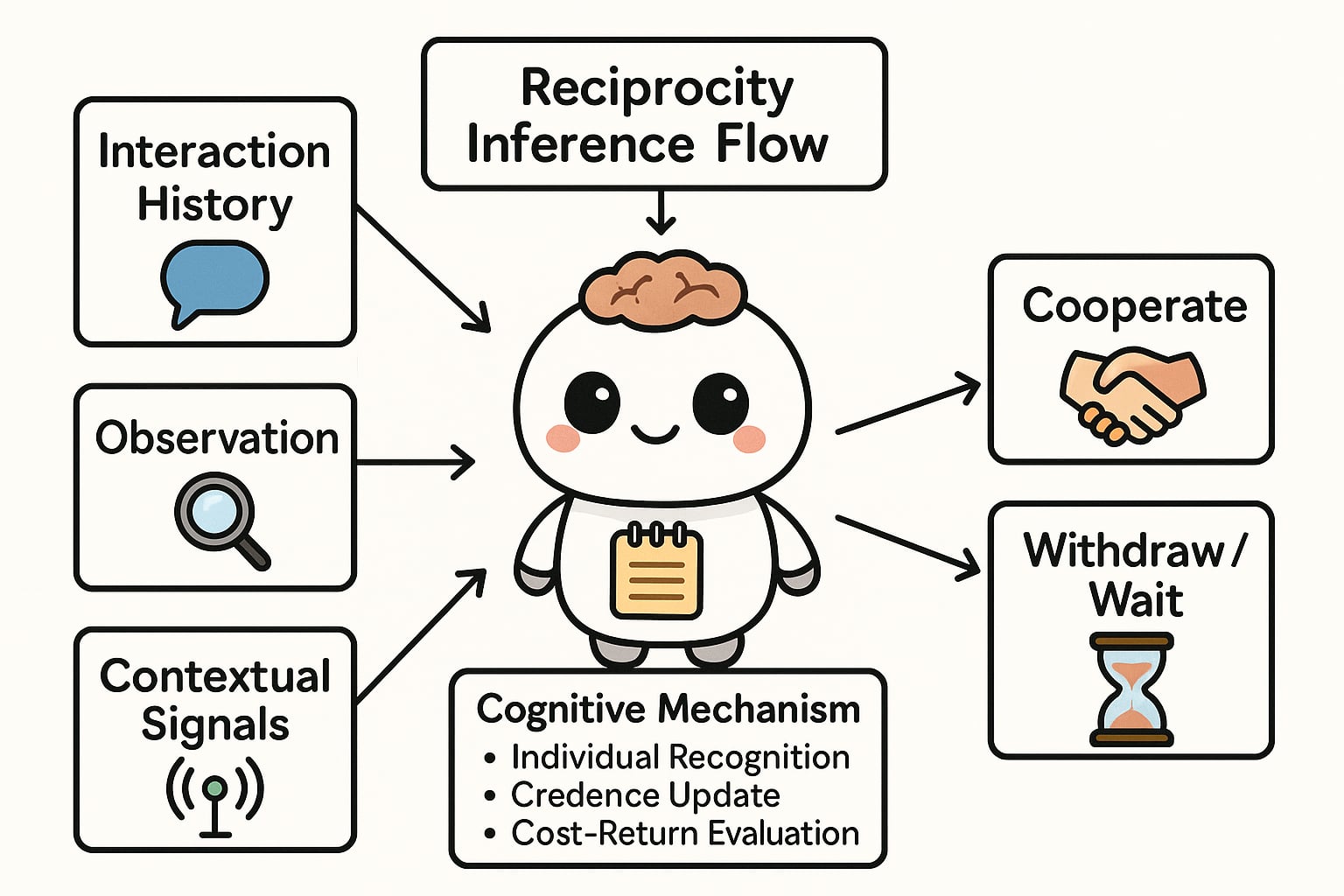}
  \caption{The agent integrates three input pathways via behaviorally grounded mechanisms—recognition, credence, and cost–return evaluation—to support dynamic decisions in reciprocal exchange.}
  \label{fig:framework}
\end{figure}

\section{Theoretical Framework: Simulateable Primitives for Economic Exchange}
\label{others}

\subsection{Individual Recognition}

\begin{definition}[Individual Recognition]
Individual recognition is the capacity to identify specific social partners over time, enabling agents to track past interactions and form expectations about future behavior. It is the prerequisite for memory-based reciprocity: without it, no past cooperation can inform future decisions.
\end{definition}

\paragraph{Empirical Support.}
Nonhuman primates show strong individual recognition. Chimpanzees and bonobos remember familiar individuals after decades and preferentially attend to former social partners \cite{lewis2023bonobos}. 

In cooperative tasks, chimpanzees adjust behavior based on partner identity and prior interactions—initiating coordination through glances, pauses, or approach behaviors—and succeed more often with tolerant, familiar partners \cite{hirata2007chimpanzees, melis2006engineering}.

In humans, this ability emerges in infancy. By 14 months, children selectively help those who previously acted prosocially \cite{dunfield2011examining}, suggesting that recognition is tied to social memory and used to guide future cooperation.

\subsection{Reciprocal Credence}
The term “trust” is widely used across disciplines, yet rarely defined with precision. It is invoked to describe moral commitment, emotional closeness, institutional reliability, and behavioral expectation—often interchangeably. This ambiguity makes it difficult to operationalize trust, especially in comparative, developmental, or artificial settings.

To address this, we introduce the concept of \textit{reciprocal credence}:

\begin{definition}[Reciprocal Credence]
Reciprocal credence is a graded, updateable belief that cooperative behavior will be reciprocated. It functions as a decision variable governing an agent’s willingness to initiate or sustain cooperation.

It arises from two temporally grounded types of inference:
\begin{itemize}
    \item \textbf{Retrospective inference}, based on memory of past prosocial behavior directed toward the self or others;
    \item \textbf{Prospective inference}, based on contextual signals that indicate future reciprocity is likely—such as social roles or reputational cues.
\end{itemize}
\end{definition}

To make these sources operational in multi-agent systems, we distinguish three canonical informational pathways:

\begin{enumerate}
    \item \textbf{Direct interaction history}, where an agent has experienced prior cooperative behavior, prosocial engagement, or affiliative signaling from a specific partner.
    \item \textbf{Third-party inference}, based on observed or reported behavior toward others—supporting third-party inference of prosocial disposition.
    \item \textbf{Role-based or contextual expectation}, where social roles, environmental incentives, or normative structures lead to default assumptions that cooperation will be returned, even in the absence of personal history.
\end{enumerate}

Together, these sources form a minimal, biologically grounded substrate for modeling reciprocal behavior in both human and artificial agents.

\paragraph{Empirical Support.} Direct positive interaction provides the most robust foundation for reciprocal credence. In chimpanzees, grooming is exchanged not only immediately but across time: dyads reciprocate over multi-day delays, suggesting memory-based calibration of cooperation \cite{gomes2009long}. They also preferentially collaborate with partners who have helped them before \cite{melis2006chimpanzees}, indicating an ability to track and evaluate individual-specific cooperative history. 

Similarly, by age three, human children avoid helping agents who previously harmed others, even if the harm was unsuccessful \cite{vaish2010young}—suggesting early sensitivity to past intent in guiding prosocial choice.

By contrast, third-party inference and role-based expectation are rarely observed in nonhuman primates and emerge only later in human development. Nonetheless, both are deeply embedded in human social life, forming the basis for reputation systems, institutional roles, and generalized cooperation beyond direct experience.

\paragraph{A note on “trust”.}
The term “trust” is often applied to systems like ChatGPT or Google, but what is described is not social trust—it is functional reliability through repeated exposure and low failure rates. These systems do not engage in reciprocal reasoning or uphold social obligations.

This semantic conflation becomes especially problematic in high-stakes domains like finance and healthcare, where “trust” is better understood as a demand for exceptional systemic robustness—not social accountability.

In finance, phrases like “trust in banks” or “trust in money” reflect belief in institutional continuity, enforcement mechanisms, and liquidity guarantees—not interpersonal confidence. What matters is structural resilience against systemic collapse.

In healthcare, “trust in doctors” or “hospital systems” similarly refers to confidence in diagnostic accuracy, institutional safeguards, and low error rates. Patients rely on these systems not because of perceived benevolence, but because failures are rare and tightly bounded.

\subsection{Cost–return sensitivity}

\begin{definition}[Cost–Return Sensitivity]
Cost–return sensitivity is the capacity to regulate cooperative behavior based on perceived payoff asymmetries over time. Rather than relying on fixed heuristics, agents adjust participation by tracking whether interactions yield sustained net benefit—enabling the avoidance of exploitation and reinforcement of beneficial exchange.
\end{definition}

\paragraph{Empirical Support.} 

In nonhuman primates, cost–return sensitivity appears in both prosocial and punitive contexts. Chimpanzees modulate grooming and cooperative behavior based on prior benefits received \cite{de1997chimpanzee}, and retaliate against theft even without immediate gain—suggesting expectations of equity and sensitivity to intentional harm \cite{jensen2007chimpanzees}.

Human infants show comparable intuitions. By 18 months, they preferentially help agents who previously acted fairly or cooperatively \cite{house2013ontogeny}, calibrating prosocial behavior not through abstract norms, but through observed patterns of giving and withholding.

Together, these findings suggest that cost–return sensitivity enables organisms to regulate cooperative investment in the absence of formal contracts—providing a cognitive substrate for reciprocity long before the emergence of markets.

\section{Simulation Pathways and Implementation}

We define three simulateable behavioral primitives—individual recognition, reciprocal credence, and cost–return sensitivity—that can serve as the minimal cognitive capacities required for scalable reciprocal behavior in multi-agent systems. Each can be approximated through lightweight memory architectures and prompt-level reasoning, without relying on explicit reinforcement learning.

\begin{itemize}
\item \textbf{Individual recognition}:
Agents should be able to differentiate between social partners and retrieve information specific to each. This may reflect accumulated positive history, social familiarity, or bonded interaction patterns—not necessarily explicit naming.

\item \textbf{Reciprocal credence}:
Agents should maintain an internal estimate of how likely a partner is to return cooperation. This estimate integrates:
(1) prior prosocial behavior directed toward the agent or observed in others, and
(2) contextual signals that make cooperative return likely.
The value is asymmetric, updateable, and sensitive to both interaction and situation.

\item \textbf{Cost–return sensitivity}:
Agents should track asymmetries in past exchanges, allowing them to adjust future cooperative investment. This mechanism supports dynamic calibration of effort and helps stabilize long-term reciprocity.
\end{itemize}

These primitives specify functional requirements rather than architectural constraints, offering a practical path toward socially grounded reciprocity in multi-agent systems.

\section{Implications and Discussion}
\label{others}

\subsection{Limitations of the Framework}

\paragraph{Comparative Methodology.}
Primate and infant studies rely on distinct paradigms: the former captures ecologically embedded behavior; the latter often uses constrained, artificial settings. This methodological gap limits fine-grained cross-species comparison of cognitive substrates.

\paragraph{Institutional Complexity.}
We do not claim that our framework alone accounts for complex institutions such as money, debt, or taxation. These might require further symbolic, cultural, and historical layers beyond minimal reciprocity mechanisms.

\paragraph{Social Scale and Enforcement.}
The current framework addresses partner-contingent reciprocity in small groups. It does not yet explain population-scale mechanisms—such as third-party punishment or distributed reputation—that enable anonymous cooperation. Bridging this gap remains a key direction for extending the model.

\subsection{Theoretical Implications}

\paragraph{Reciprocity as the Foundation of Social Structure.}
The term \textit{cooperation} is frequently used as a catch-all for prosocial behavior, but its vagueness makes it difficult to model or implement. Without a clear cognitive basis, many systems rely on artificial payoff matrices, fixed strategies, or exogenous incentives—mechanisms that simulate outcomes but not the behavioral processes behind them.

In contrast, reciprocity offers a cognitively grounded and behaviorally structured alternative. Rather than relying on externally defined payoffs, it emerges from simple agent-level mechanisms that support dynamic, partner-contingent decision-making. This makes reciprocity tractable, simulateable, and aligned with real-world behavior.

More importantly, reciprocity underlies many core functions of modern society. From social favors and mentorships to business alliances and venture capital, cooperation often depends on asymmetric, memory-based, and deferred exchanges. Even scientific systems—such as peer review, citation, and referrals—operate through implicit reciprocity, not formal enforcement.

We argue that reciprocity—not abstract cooperation—is the minimal behavioral substrate of social structure. It enables agents to build conditional expectations, regulate ongoing interactions, and scale bilateral exchange into stable institutional patterns. Far from being a subset of cooperation, reciprocity is the mechanism that makes cooperation sustainable and socially extensible in the first place.

\paragraph{From Institutions to Cognition.}
Our work reorients the study of exchange from an institutionalist or symbolic perspective toward a cognitive-behavioral foundation. Rather than treating markets, money, or debt as cultural inventions that enabled exchange, we argue that exchange itself is grounded in prior cognitive and social capacities—capacities that predate and scaffold institutionalization.

\paragraph{Rethinking Trust as System Robustness.}
The term “trust” is often invoked in discussions of economic exchange, yet it bundles together distinct ideas—from interpersonal reliability to confidence in systems. We argue that this vagueness limits its explanatory value.

To address this, we introduce \textit{reciprocal credence} as a graded, behaviorally grounded expectation—specific to reciprocal exchange—that governs whether an agent initiates or maintains cooperation. It depends not on social bonding, but on observable interaction history and contextual inference.

In contrast, large-scale “trust” in systems like money or platforms is better understood as confidence in system robustness—not interpersonal accountability or relational bonding. This reframing clarifies how large-scale exchange systems can emerge without relying on social-bond-based trust.

\paragraph{Cross-Species Foundations of Exchange.}
By aligning primate cooperation studies with human developmental data and ethnographic accounts of non-monetary exchange, we provide a cross-species bridge for modeling the origins of economic behavior. This integrative view challenges the idea that complex exchange systems are uniquely human, and instead locates their roots in broader social cognition.

\paragraph{Unifying Behavioral Economics through Cognitive Foundations.}
If economic behavior is an extension of biologically grounded reciprocity, then so-called “biases” identified in behavioral economics—such as anchoring bias, loss aversion are not anomalies—they are direct extensions of biological traits.

\subsection{Future Work}

\paragraph{Operationalizing Economic Emergence.}
A critical next step is to embed our proposed primitives—\textit{individual recognition}, \textit{reciprocal credence}, and \textit{cost–return sensitivity}—as explicit modules in multi-agent simulations. Future experiments should systematically test how memory constraints, scalar reciprocity estimates, and payoff asymmetries influence the emergence of sustained cooperation and reciprocal exchange structures.

\paragraph{Institution Formation via Behavioral Mechanisms.}
Further simulations should explore whether agents equipped with these minimal cognitive primitives spontaneously develop proto-institutions such as symbolic debt, token systems, or role-based cooperation. Advances in embodied simulators (e.g., VirtualHome \cite{puig2018virtualhome}, AutoGen \cite{wu2023autogen}) offer opportunities to bridge cognitive theories of exchange with practical implementations of scalable social structures.

\paragraph{Reciprocity Among Unfamiliar Agents.}
Finally, investigating reciprocal exchange among unfamiliar agents—without prior recognition or shared interaction history—represents an important frontier. Understanding the cognitive scaffolds enabling such interactions could significantly advance our grasp of decentralized cooperation mechanisms.

\section{Conclusion}
Contrary to the classical narrative of barter, converging evidence from early human societies suggests that economic life began with reciprocity—long-term, partner-contingent exchange embedded in social relationships. This behavioral substrate enabled communities to circulate resources, maintain cohesion, and sustain cooperation well before markets or institutions emerged.

Yet despite its foundational role in early economies, reciprocity remains under-formalized in both economic theory and computational modeling. We propose a cognitively grounded framework that identifies three simulateable mechanisms—individual recognition, reciprocal credence, and cost–return sensitivity—as the minimal substrate for scalable exchange.

By grounding abstract terms like “trust” and “cooperation” in observable mechanisms, this approach shifts economic modeling from top-down design to bottom-up simulation. These primitives not only clarify the origins of exchange, but also provide a tractable substrate for implementing reciprocal behavior in multi-agent systems.





\section*{Impact Statement}
This paper introduces a cognitively grounded framework for simulating the behavioral foundations of economic exchange, based on biologically plausible primitives. By shifting the focus from top-down incentives to bottom-up social cognition, this approach enables more realistic and interpretable models of reciprocal cooperation. We anticipate that this framework could significantly improve agent-based economic simulations, particularly in multi-agent systems concerned with trust, fairness, and coordination. While the work is theoretical and does not involve deployment or data collection, we are not aware of any negative societal or ethical risks associated with this research.

\nocite{*}
\bibliography{example_paper}
\bibliographystyle{icml2025}

\newpage
\appendix
\onecolumn



\end{document}